\documentclass[journal]{IEEEtran}
\usepackage{booktabs}      % <-- Place it here
\usepackage{amsmath}       % If you need math
\usepackage{graphicx}      % For figures
\usepackage{float}
\usepackage{orcidlink}

\hypersetup{
    hidelinks
}

\ifCLASSINFOpdf
\else
\fi
\begin{document}
\title{Reconfigurable Nonlinear Photonic Networks for In-Situ Learning and Memory Formation via Driven–Dissipative Dynamics}

%\author{Isaac Yorke}
\author{Isaac Yorke, \orcidlink{0000-0002-1645-6470}\\
 Department of Engineering and Architecture, \\University of Parma, Parco Area delle Scienze 181/A I-43124 Parma, Italy\\
e-mail: isaac.yorke@unipr.it  or  yorkeisaac034@gmail.com}

%\markboth{This work is submitted to IEEE Photonics Journal}
\markboth{This manuscript has been submitted to \textit{Neuromorphic Computing and Engineering}}
{Shell \MakeLowercase{\textit{et al.}}: Bare Demo of IEEEtran.cls for IEEE Journals}



\maketitle

\begin{abstract}
Photonic neuromorphic computing offers a promising route to overcoming the limitations of conventional von Neumann architectures by exploiting the high bandwidth, low latency, and massive parallelism of optical systems. However, most existing implementations rely on fixed dynamical substrates such as classic reservoir computing, where learning is restricted to external readout layers and memory is limited to transient fading effects. This work proposes a Reconfigurable Nonlinear Photonic Decision Network (RNPDN), a physically grounded neuromorphic framework in which computation, memory, and learning emerge directly from driven–dissipative dynamics. Through numerical simulations, this work demonstrates the simultaneous realization of key properties: local physical learning rules enabling adaptive state evolution, a tunable stability–plasticity tradeoff governed by decay and hysteresis mechanisms, controlled memory formation and erasure via bistable photonic states, fading memory, in-situ learning, and hardware-faithful nonlinear dynamics incorporating saturation and dissipation. In contrast to conventional approaches, the proposed system enables intrinsic adaptation within the physical layer while supporting both transient and persistent memory. These results establish a unified framework for adaptive photonic information processing and provide a pathway toward scalable and energy-efficient neuromorphic photonic hardware.
\end{abstract}


% \fi
\IEEEpeerreviewmaketitle

\section{Introduction}
\IEEEPARstart{I}{n} recent years, photonic neuromorphic computing has emerged as a promising paradigm to overcome the fundamental limitations of conventional von Neumann architectures, particularly in terms of data movement, latency, and energy consumption. Photonic implementations leverage the intrinsic advantages of light; such as high bandwidth, parallelism, and low dissipation to enable ultrafast and energy-efficient information processing. Within this domain, reservoir computing (RC) has attracted significant attention due to its simplified training requirements, where only the readout layer is optimized while the internal dynamics remain fixed \cite{borghi2021reservoir}. A wide range of photonic neuromorphic systems have been proposed and demonstrated based on nonlinear dynamical elements, including semiconductor lasers, microcavities, and microring resonators. For instance, microring-based reservoir computing platforms have shown the ability to perform both analog and digital processing tasks by exploiting nonlinear optical dynamics and time multiplexing \cite{borghi2021reservoir}. More advanced architectures, such as multi-layer microring reservoirs, have been introduced to enhance both nonlinearity and temporal memory, significantly improving performance on benchmark time-series prediction tasks \cite{dong2026deep}. Similarly, integrated photonic neural networks based on tunable microring resonators have demonstrated efficient pattern recognition capabilities, highlighting the potential of silicon photonics for scalable neuromorphic hardware \cite{zhang2025chip}. In parallel, photonic reservoir systems based on semiconductor lasers and delay dynamics have been explored for high-speed temporal processing, although their performance is often constrained by limited node scalability and feedback architectures \cite{guo2024photonic}. Despite these advances, existing photonic neuromorphic systems exhibit several fundamental limitations. First, most implementations rely on fixed internal dynamics, with learning restricted to an external readout layer, thereby limiting the system’s ability to adapt or evolve in response to environmental stimuli \cite{borghi2021reservoir}. Second, the dominant computational paradigm is based on fading memory, where past inputs influence the system state only transiently, leading to limited long-term memory retention. Although recent efforts have explored extended memory through coupled resonator networks and multistability, these approaches remain constrained in their ability to realize controlled memory formation and erasure \cite{lugnan2025reservoir}. Finally, the absence of intrinsic physical learning mechanisms such as local, device-level adaptation prevents these systems from achieving true in-situ learning, a key requirement for scalable neuromorphic hardware. 
This work introduces a Reconfigurable Nonlinear Photonic Decision Network (RNPDN), a photonic neuromorphic computing framework in which computation, memory, and learning emerge directly from the underlying driven–dissipative physical dynamics of the system. Unlike conventional reservoir computing approaches, the proposed system illustrates via simulations: (i) physical learning rules, where adaptation arises from local, device-consistent updates rather than externally imposed optimization; (ii) a tunable stability–plasticity tradeoff, governed by intrinsic decay and hysteresis mechanisms; (iii) memory formation and erasure enabled by bistable photonic states, allowing both short-term (fading) and long-term memory coexistence; (iv) In-situ learning and (v) hardware-faithful nonlinear dynamics, explicitly incorporating physical constraints such as saturation and dissipation. By unifying these features within a single dynamical framework, the RNPDN becomes anchored in physics and moves beyond the conventional separation between computation and learning in photonic systems. This work illustrates the the feasibility of a possible new paradigm for adaptive photonic intelligence, where the system is not only capable of processing temporal information but also of self-modifying its internal state in response to stimuli. This work also establishes the conceptual viability of a pathway toward fully integrated, energy-efficient, and physically realizable neuromorphic photonic processors with enhanced computational capabilities. Previous studies have extensively investigated individual physical mechanisms underlying photonic information processing, including optical bistability and hysteresis in nonlinear resonators \cite{gibbs2012optical}, saturation effects, fading memory in photonic reservoir computing \cite{vandoorne2014experimental}, and driven–dissipative nonlinear dynamics \cite{herr2016dissipative}. The novelty of the present work does not lie in these individual mechanisms themselves, but to illustrate their integration within a reconfigurable photonic decision network that enables adaptive computation through local driven–dissipative dynamics.

\section{Physical Model and System Description}
To establish a physically grounded framework for photonic neuromorphic computation, this section introduces the dynamical model underlying the proposed RNPDN. The system is formulated within a driven–dissipative paradigm, where the interplay between external excitation, intrinsic nonlinear response, and dissipation governs the evolution of the internal states. Such dynamics are characteristic of a wide class of integrated photonic platforms, enabling a direct connection between the mathematical description and experimentally realizable devices.
The model is expressed through the laser rate equations, which is a typical example of driven–dissipative systems and also captures key physical mechanisms, including decay, saturation, pump power, state-dependent feedback, hysteresis, etc \cite{erneux2010laser}. These mechanisms give rise to multistability and temporal memory, which form the basis for the observed computational properties. In addition to presenting the governing equations, their physical interpretation is discussed to clarify the role of each term in shaping the system dynamics.
Finally, the abstract model is mapped onto representative photonic implementations, such as Microring Resonator-based platforms, where nonlinear optical effects and cavity dynamics naturally support driven–dissipative behaviour. This mapping highlights the feasibility of realizing the proposed framework in integrated photonic hardware and provides a bridge between theoretical modeling and practical implementation.

\subsection{Driven–Dissipative equations }
Driven–dissipative systems provide a natural framework for describing the dynamics of nonlinear photonic platforms \cite{haken1973laser, brunner2013parallel}. In such systems, the evolution of the state variables is governed by the continuous interplay between external driving, intrinsic nonlinear interactions, and energy dissipation \cite{lugiato1987spatial}. This class of dynamics has been extensively studied in the context of optical cavities, semiconductor lasers, and integrated photonic resonators, where it gives rise to rich phenomena including multistability, hysteresis, and self-organization \cite{erneux2010laser, haken1973laser, lugiato1987spatial}. Mathematically, driven–dissipative behaviour is typically described by nonlinear differential equations of the form:   
\begin{equation}
\frac{dS}{dt} = F(S,I) - \gamma S
\end{equation}
where $S$ represents the system state, $I$ denotes the external input or driving term, $F(.)$ captures nonlinear response mechanisms such as saturation or feedback, and $\gamma$ accounts for dissipation \cite{erneux2010laser, lugiato1987spatial}. If the system is externally driven (e.g; by pump power, optical feedback, etc), then a  nonequilibrium steady state can be reached, and thus; $\frac{dS}{dt} = 0 $, hence; $F(S,I) = \gamma S $, that is; $Driving = Dissipation$. This means, at the nonequilibrium steady state, the external driving and dissipative processes balance each other, resulting in a stationary regime where the temporal derivative of the state variables approaches zero \cite{lugiato1987spatial}. Laser rate equations describe a driven–dissipative system in which external pumping continuously injects energy, cavity and material losses continuously remove energy, and the steady laser output arises from their nonlinear balance. The coexistence of driving and dissipation enables the emergence of stable nonequilibrium states, which can be harnessed for information storage and processing. In the context of photonic neuromorphic systems, these dynamics provide a physically grounded mechanism for implementing memory, nonlinearity, and adaptive behaviour within a unified framework \cite{ramanathan1945introduction}.

\subsection{The Learning Model}
In order to anchor the proposed system in physics, learning in the RNPDN is governed by a local, analog update rule in which each weight evolves according to its instantaneous activation and a scalar reward signal. This form mirrors driven-dissipative photonic dynamics and requires no global error propagation or centralized control, making it compatible with integrated photonic hardware. The model captures the competition between passive relaxation (loss) and external optical driving, which constitutes the fundamental dynamical mechanism underlying driven–dissipative nonlinear photonic systems. \cite{lugiato1987spatial, herr2016dissipative}. Within the proposed RNPDN framework, the external optical driving is represented conceptually by a local reinforcement term that modifies the effective optical weight during each decision cycle. Suppose the effective optical weight $w_i(t)$ represents the transmission or phase state of a tunable photonic element (for example, a thermo-optic phase shifter, carrier-injection modulator, or nonlinear resonator). A simple physical model for its evolution is $\frac{dw_i}{dt} = -\Gamma w_i + P_i(t)$, where $w_i=$ optical state (effective weight), $\Gamma=$ passive relaxation (optical loss, thermal relaxation, carrier recombination), and $P_i=$ external optical excitation or control. This equation represents a generic first-order relaxation model widely used to describe the evolution of effective optical states in driven photonic systems, where the dynamics arise from the competition between intrinsic relaxation and external excitation \cite{lugiato1987spatial, haken1985light}. Let $P_i(t)=\eta \Delta w_i(t)$, where $\eta= $ learning efficiency(optical pumping efficiency) and $\Delta w_i=$ reward/optical feedback(locally generated reinforcement signal). This gives $\frac{dw_i}{dt} = -\Gamma w_i(t) + \eta \Delta w_i(t)$. Assuming the photonic system is updated once every decision cycle of duration $\Delta t$, the continuous-time dynamics can be discretized using the Forward Euler  first-order approximation method. In the Forward Euler approximation, $\frac{dw_i}{dt} \approx \frac{w_i(t+\Delta t)-w_i(t)}{\Delta t}$ \cite{chapra2011numerical}, thus;  $\frac{w_i(t+\Delta t)-w_i(t)}{\Delta t}=-\Gamma w_i + \eta \Delta w_i(t)$ $\Rightarrow$ $w_i(t+\Delta t) = \Delta t(-\Gamma w_i(t) + \eta \Delta w_i(t))+ w_i(t)$. $\Rightarrow$ $w_i(t+\Delta t)=w_i(t)- \Gamma \Delta t w_i(t)+\Delta t \eta \Delta w_i(t))$. Let $\gamma = \Gamma \Delta t$ and $\Delta t=1$ gives the learning model of the proposed framework as
\begin{equation}
w_i(t+1) = (1-\gamma) w_i(t) + \eta \, \Delta w_i(t)
\label{learning model}
\end{equation}
For $\Delta w_i(t) \approx 0$ (no strong reward), $\frac{dw_i}{dt} = -\Gamma w_i$, whose solution is $w_i(t)=w_i(0)e^{-\Gamma t}$, and sampling every update interval gives   
\begin{equation}
w_i(n+1) \approx (1-\gamma) w_i(n) 
\label{exponential decay eqn}
\end{equation}
Thus; an exponential decay equation \cite{elaydi2005introduction}. The physical meaning is that, the photonic memory slowly leaks due to carrier recombination, or optical loss \cite{lugiato1987spatial, xu2005micrometre}.

\subsubsection{Mapping RNPDN Directly onto Physical Hardware}
It is established that the learning model in Equation \ref{learning model} is based on Driven-Dissipation system. This means that the physical features being dealt with are phase shift, optical feedback, pump power strength, absorption/leakage, etc. It is known that these physical features are subject to decay, hysteresis, and saturation. What this means is that learning is intrinsically nonlinear, history dependent, and bounded. There are some photonic devices such as phase changes materials, carrier-based modulators, ring resonators, etc that exhibit these types of behaviours \cite{ramanathan1945introduction, lugnan2022rigorous, lugnan2023silicon}. This means that the RNPDN learning model can be directly mapped unto these physical hardware. For instance, in microring resonators, intensity-dependent refractive index changes arising from Kerr and thermo-optic nonlinearities can shift the cavity resonance, resulting in bistable and hysteretic transmission characteristics \cite{ja1993kerr, zhang2013multibistability}. The RNPDN departs fundamentally from traditional machine learning systems in both representation and learning mechanism. In contrast to gradient-based optimization over continuous loss functions, learning in the RNPDN is driven by local, reward-modulated weight updates coupled to bistable hysteretic state transitions. It also departs from classical reservoir computing system where learning takes place only at the outer layer. This is a concept to implement a “beyond gradient-based photonic learning” framework. Having anchored the learning model (Equation \ref{learning model}) in physics, the physical learning rule demonstration is established.
\section{Simulation Framework}
To illustrate the principle of programmable multi-state photonic memory, this work employs a phenomenological model where computational parameters of the learning model in Equation \ref{learning model}; weights, decay, reward, learning rate are mapped to physical photonic quantities; phase shift, loss, pump power, optical feedback. The simulations presented in this work are intended as an early proof-of-concept demonstration rather than a quantitative model of a specific photonic device. The parameter values were selected to produce stable driven–dissipative dynamics within the normalized phenomenological framework introduced in the derivation of the learning model and are representative of the qualitative operating regime rather than any particular hardware implementation. Accordingly, the results should be interpreted as demonstrating the feasibility of the proposed learning mechanism and its emergent dynamical behavior. Future work will calibrate the model using experimentally measured device parameters and perform quantitative comparisons with physical photonic platforms. Finite photon lifetimes, thermal feedback, carrier dynamics, and cavity-mediated interactions are all natural memory reservoirs in photonic systems. Consequently, memory is an intrinsic property of photonics \cite{biasi2026phase}. Based on this notion, rather than starting from a conventional neural network abstraction or gradient-based optimisation, the RNPDN is grounded in a competitive, game-based interaction between two players (agents). The game serves not merely as a benchmark, but as a structured environment in which learning must occur sequentially, under delayed reward, partial observability, and adversarial pressure. These properties naturally expose the limitations of purely algorithmic learning models while providing a rigorous testbed for physically constrained learning dynamics. Detailed description of the game (known in Ghana as Oware) can be found in \cite{bayeck2018review}. The reward in the learning model is then obtained through the player (agent) score.
To maintain asymmetric photonic hysteresis and bistability which induces memory in the system, some hysteresis parameter values were chosen for the device physics. The control of such hysteresis parameters in a microring resonator, and their dependence on system variables like detuning, has been previously analyzed  based on studies from \cite{biasi2026phase, zhang2026multiscale, bahrampour2008all, lentine1988symmetric}. These values are labeled as: THETA\_SET\_POS, THETA\_RESET\_POS, THETA\_SET\_NEG, and THETA\_RESET\_NEG. The values of $\eta$ and $\gamma$ were chosen based on guidance from  \cite{saxena2024circlez}. Even though the ranges of these two parameter values were chosen to be the same, their optimal values were different. Table \ref{tab:table_for_simulation_values}
shows the values that were set for the simulation.

\begin{table}[h]
\centering
\caption{Table of values for simulation.}
\label{tab:table_for_simulation_values}
\begin{tabular}{cc}
\toprule
\textbf{Parameter} & \textbf{Values} \\
\midrule
$Decay (\gamma)$ & $(10^{-5} - 10^{-2})$ \\
$Learning\ efficiency (\eta)$ & $(10^{-5} - 10^{-2})$ \\
$Weights\ (w)$ & {random number between: 0 and 1} \\
\texttt{THETA\_SET\_POS} & 1.8 \\
\texttt{THETA\_RESET\_POS} & 0.8 \\
\texttt{THETA\_SET\_NEG} & 1.2 \\
\texttt{THETA\_RESET\_NEG} & 0.6 \\
\bottomrule
\end{tabular}
\end{table}

 

\section{Results and Analysis}
This section presents a systematic evaluation of the dynamical behaviour and computational properties of the proposed RNPDN through numerical simulations. The analysis is structured to isolate and validate the key physical mechanisms underlying the system’s operation. First, the intrinsic nonlinear dynamics governing state evolution are examined to establish the foundation for complex behaviour. Next, the emergence of bistability and its role in enabling robust memory formation and controlled state transitions are then demonstrated. The characterization of the system’s fading memory response, highlighting its ability to retain and process temporal information is also demonstrated. Building on these results, the stability–plasticity tradeoff is investigated to illustrate how the system balances adaptability with retention under varying conditions. The impact of hardware-faithful constraints, including saturation and dissipation, is also analyzed to emphasize the importance of physical realism in achieving stable and reliable operation. The potential scalability of the system to multiple independent channels or ports is illustrated and finally, the capability of the system to perform in-situ learning is demonstrated. Together, these results provide preliminary support for the proposed framework and demonstrate its potential as a physically inspired approach to photonic neuromorphic computing.

%\IEEEpubidadjcol

\subsection{Nonlinear Dynamics}
\begin{figure}[!htbp]
    \centering
    \includegraphics[width=0.45\textwidth]{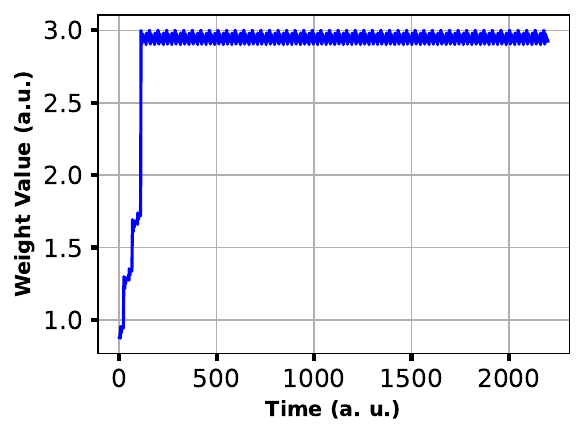}
    \caption{Time evolution of the weight trajectory after removing nonlinear effects. The weight rapidly saturates at 3 (a.u.) and exhibits no further dynamics, confirming the absence of bistability and memory.}
    \label{fig:nonlinear_dynamics}
\end{figure}

In photonics and all physical systems, memory and computation require nonlinearity. Without nonlinearity, no information processing is possible, and memory is a form of information processing \cite{schulte2023refined, horn1999importance, suarez2021network}. In dynamical systems, there exists a direct trade-off between a system's capacity for memory and its degree of nonlinearity \cite{schulte2023refined}. Without nonlinear elements, a system's dynamics remain linear and cannot support the multiple stable states or persistent activity patterns necessary to store information. For instance, in neural network models, nonlinearities in neuronal transfer functions are essential for stabilizing persistent activity during working memory tasks \cite{suarez2021network}. To demonstrate the relevance of nonlinearity in the RNPDN, the nonlinear component of the system was put off and the time evolution of the weight trajectory was simulated. Figure~\ref{fig:nonlinear_dynamics} shows the time evolution of the weight trajectory when nonlinear  effect is removed. Removing the nonlinear dynamics from the model collapses the system to a linear saturating response. This is not a memory, but a saturation curve. Once the weight hits maximum value (3 (a. u.) in this case), it becomes a flat line with no dynamics, no switching, and no state retention beyond a single value. This confirms that nonlinearity is essential for generating the bistability and hysteresis required for photonic memory. Without nonlinear effects, the system cannot support multiple stable states and therefore cannot function as a memory element. 

In this system, time is obtained through the relation $t=n*\Delta t$. Where the variable $n$ is the integer cycle index, and $\Delta t$ is the fixed time interval between cycles \cite{oppenheim1999discrete, proakis2007digital}. Since the system is simulation-based, $\Delta t$ is chosen to be $1$ and therefore $t= \text{cycle index}$. Time is therefore represented by the discrete cycle index, corresponding to the iterative evolution of the system.

\subsection{Memory Formation and Bistability}
\begin{figure}[!htbp]
    \centering
    \includegraphics[width=0.45\textwidth]{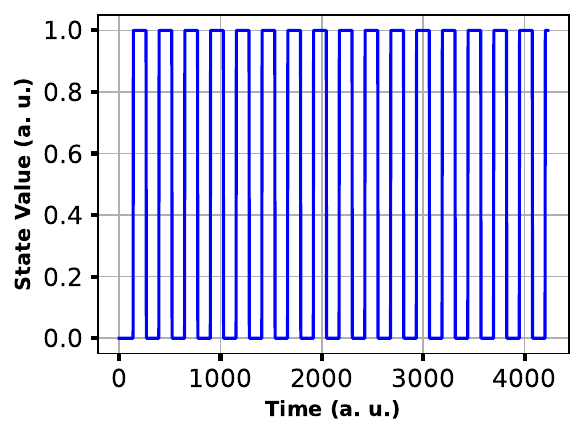}
    \caption{Bistable switching behaviour of the photonic memory state. The state toggles between stable low (0) and high (1) states, remains saturated in each state for multiple times, and then toggles back. This demonstrates reproducible photonic switch operation with clear set and reset transitions, analogous to recently demonstrated programmable photonic latches \cite{ashtiani2025programmable}.}
    \label{fig:state_trajectory}
\end{figure}

Memory formation in photonic neuromorphic systems requires the existence of stable states that can persist in the absence of continuous input. In the proposed RNPDN framework, when the nonlinear component is switched on, this capability arises from bistable dynamics, where the system can reside in one of two distinct stable states depending on its input history. Such behaviour is a hallmark of nonlinear driven–dissipative systems and has been extensively studied in optical cavities and resonator-based platforms, where it enables optical memory and switching functionalities \cite{gibbs2012optical}. In this subsection, the emergence of bistability is analyzed through the system’s state and weight trajectories. The results demonstrate how transitions between stable states enable reliable memory formation, while the inherent nonlinearity ensures robustness against noise and fluctuations. This establishes bistability and hysteresis as key physical mechanisms underpinning persistent memory in the proposed architecture.
Figure~\ref{fig:state_trajectory} shows the time evolution of the state variable for the system, which demonstrates bistable toggling between two states (0 and 1) across multiple times. The state alternates between two stable values (0 and 1), remaining saturated in each state for several times before toggling. The flat plateaus indicate robust state retention, while the sharp transitions demonstrate fast switching between memory states \cite{ashtiani2025programmable}.

\begin{figure}[!htbp]
    \centering
    \includegraphics[width=0.45\textwidth]{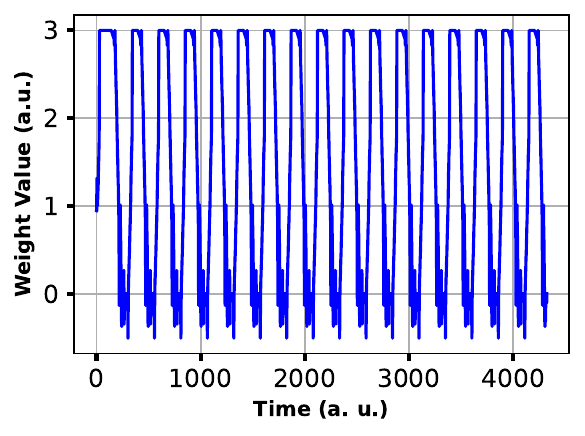}
    \caption{Time evolution of the photonic weight trajectory, which reveals three distinct dynamical regimes. The hysteresis regime which stretches from $0 (a. u.)$ to $1 (a. u.)$ portion of the weight values, reflecting the system's plastic regime. The relaxation regime stretching from maximum weight ($3 (a. u.)$) and descend to $~1 (a. u. )$, displaying a gradual decay. Lastly, the bistable memory regime at maximum weight ($3 (a. u.)$), which shows the weight saturating at a stable plateau value, confirming long-term memory retention. \cite{ashtiani2025programmable}.}
    \label{fig:weight_trajectory}
\end{figure}

In this system, memory equals persistence of a decision state due to hysteresis. A memory is formed when RNPDN weights remain in a stable hysteretic branch over consecutive times despite zero or fluctuating reward signals. Figure~\ref{fig:weight_trajectory} shows the time evolution of the photonic weight trajectory, revealing three distinct dynamical regimes. First, the hysteresis regime which stretches from $0 (a. u.)$ to $1 (a. u.)$ portion of the weight values, reflecting the system's plastic regime. Second, the relaxation regime stretching from maximum weight ($3 (a. u.)$) and descend to $~1 (a. u. )$, displaying a gradual decay, which is attributed to the inherent erasure or forgetting mechanism, a feature essential for fading memory and temporal information processing. Finally, the bistable memory regime at maximum weight ($3 (a. u.)$), which shows the weight saturating at a stable plateau value, confirming long-term memory retention. Table \ref{tab:clarify memory formation plot} shows the interpretation of the oscillations and plateau regions of Figure~\ref{fig:weight_trajectory}.
The plot shows that the system exhibits fast transitions (learning/plasticity), followed by long plateaus near maximum weight (Bistable memory regime/memory Formation), then relaxes (decay) to enter the hysteretic regime and the cycle repeats. This matches physical hysteretic memory, not algorithmic storage. It is not a simple saturation because, weights leave the plateau, then return, small fluctuations do not erase the plateau and transitions require threshold crossings. Once weights enter a hysteretic branch (maximum weight), learning updates no longer produce changes, the system retains its internal state for a relatively long time. This persistence is the memory. This is exactly how Phase-change materials \cite{pries2019switching}, Microring bistability \cite{suresh2025hysteresis} and Carrier-induced nonlinearities \cite{nikitin2022investigation} store information.

\begin{table}[h]
\centering
\caption{Table which clarifies memory formation in Figure~\ref{fig:weight_trajectory}.}
\label{tab:clarify memory formation plot}
\begin{tabular}{cc}
\toprule
\textbf{Region} & \textbf{Interpretation} \\
\midrule
Rapid oscillations & Plastic learning regime \\
Entry into plateau & Memory formation event \\
Long flat region & Stored memory \\
Exit from plateau & Memory erasure / rewriting \\
\bottomrule
\end{tabular}
\end{table}

\subsection{Fading memory}
Fading memory is a fundamental property of dynamical systems used for temporal information processing, referring to the ability of the system to retain a history of past inputs with progressively diminishing influence over time. In photonic neuromorphic systems, this property enables the encoding of temporal correlations without the need for explicit storage elements \cite{boyd2003fading, duport2012all}. In the proposed RNPDN framework, fading memory naturally arises from the driven–dissipative dynamics, where the balance between input excitation and intrinsic loss governs the persistence of system states. In this subsection, the temporal response of the system based on decay inputs is analyzed to quantify the decay characteristics. 
Equation \ref{exponential decay eqn} already reveals the exponential decay nature of the system. To estimate the fading characteristics of the system, the relation:
\begin{equation}
w(t) \approx e^{-\gamma t}
\label{eq:fading memory eqn}
\end{equation}
is employed, where $w(t)=\text{weights}$, $\gamma = \text{Decay}$ and $t=\text{Time}$  \cite{boyce2021elementary, neftci2019surrogate}.

\begin{figure}[!htbp]
    \centering
    \includegraphics[width=0.45\textwidth]{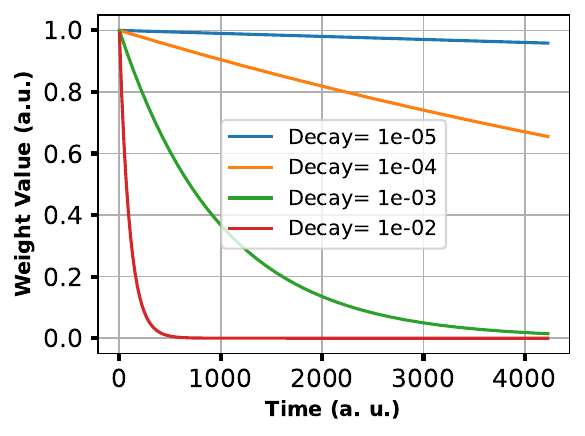}
    \caption{Smaller decay values (e.g., $1\times10^{-5}$) exhibit slower decay, corresponding to longer fading memory timescales, while larger decay values (e.g., $1\times10^{-2}$) produce rapid decay, corresponding to shorter memory. This demonstrates the tunability of the system's temporal processing capability.}
    \label{fig:fading_memory}
\end{figure}

Figure~\ref{fig:fading_memory} characterizes the fading memory of the photonic system by plotting the weight evolution under exponential decay, $W(t) \approx e^{-\gamma t}$, for different values of decay parameter $\gamma$. The decay parameter controls the memory timescale: smaller $\gamma$ (e.g., $1\times10^{-5}$) yields a slowly decaying weight that retains information over long temporal windows, while larger $\gamma$ (e.g., $1\times10^{-2}$) yields rapid decay, effectively erasing past inputs quickly. This tunable fading memory is essential for temporal information processing tasks, where the system must balance retention of history with sensitivity to recent inputs \cite{carroll2022optimizing, vrugt2024introduction}.

\begin{figure}[!htbp]
    \centering
    \includegraphics[width=0.45\textwidth]{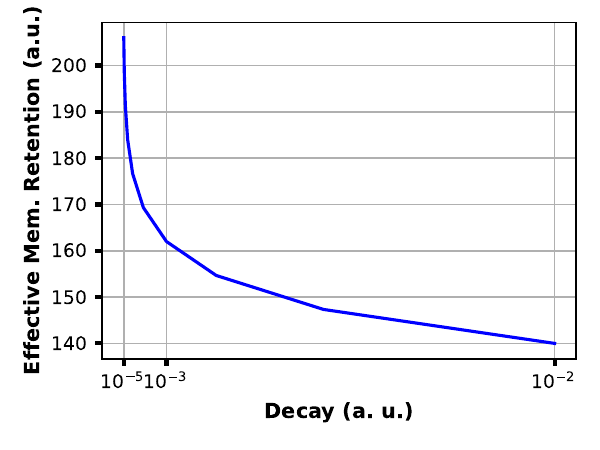}
    \caption{Quantitative demonstration of effective system memory retention vs decay, which reveals that information persistence decreases as the decay parameter increases. This confirms that past inputs have diminishing influence over time and that the memory depth of the system is physically controllable, which is a key property of fading memory in dynamical systems. \cite{jaeger2001echo, du2017reservoir, hermans2010memory}}
    \label{fig:memory lifetime vs decay}
\end{figure}

The effective system memory retention, which is defined as the maximum number of consecutive update cycles during which the sign of a learned weight remains unchanged after thresholding was also investigated. It should be noted that the RNPDN memory lifetime defined here is a computational measure of information persistence and should not be interpreted as the intrinsic exponential relaxation time associated with the isolated weight dynamics described by equation \ref{eq:fading memory eqn}. Rather, it quantifies the persistence of stable decision states arising from the complete driven–dissipative learning dynamics of the network. For the decay values $[10^{-5}, 10^{-4}, 10^{-3}, 10^{-2}]$, the corresponding  effective system memory retention values obtained from the RNPDN were; $[206, 194, 142, 140]$. A logarithmic sweep of 10 values was generated for the decay  values. Similarly, a linear sweep of 10 values was generated for the system memory lifetime values. These 10 values were obtained by linear interpolation between the specified bounds. The relationship between effective system memory retention and the decay parameter is shown in Figure~\ref{fig:memory lifetime vs decay}. As the decay coefficient increases, the effective memory retention decreases monotonically, confirming the fading memory behaviour of the system. Figure~\ref{fig:memory lifetime vs decay} shows that fading memory exists, memory is not fixed, it is tunable and it is controlled by a physical parameter such as decay. Importantly, this result demonstrates that the memory depth of the RNPDN can be continuously tuned via the decay parameter, providing a controllable tradeoff between temporal integration and responsiveness. Such tunability is essential for tasks involving sequential decision-making, where the system must balance long-term information retention with adaptability to new inputs \cite{jaeger2001echo, du2017reservoir, hermans2010memory}.

\subsection{Stability-Plasticity Tradeoff}
\begin{figure}[!htbp]
    \centering
    \includegraphics[width=0.45\textwidth]{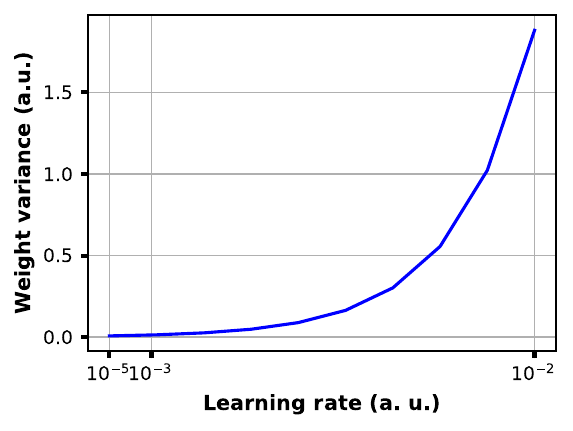}
    \caption{The stability-plasticity tradeoff of the system. Importantly, this result demonstrates that the RNPDN framework allows continuous tuning between stable and plastic regimes through a physically meaningful parameter, enabling the system to balance memory retention and adaptability depending on the task requirements. This behaviour is consistent with the stability–plasticity dilemma, where excessive plasticity leads to instability, while excessive stability limits learning capacity \cite{mermillod2013stability}.}
    \label{fig:weight variance}
\end{figure}

In photonic hardware and biological learning systems, there is a fundamental tradeoff; Plasticity, that is, the ability to learn new patterns and Stability, that is the ability to retain learned patterns (memory) \cite{mermillod2013stability}. Systems must simultaneously acquire new knowledge and retain old knowledge, and this balance is central to continual learning \cite{feng2025balancing}. Learning modifies synapses, which can induce interference with stored memory and this creates an inherent tradeoff between plasticity and stability \cite{natrajan2025stability}. The system has all the ingredients that can demonstrate the stability-plasticity tradeoff. These are learning; which emerges from reward-based updates, decay; which is due to photonic leakage, and bistable memory due to hysteresis. The concept of weight variance was used to quantify the stability-plasticity tradeoff of the system. The weight variance is defined as the statistical measure of the spread or fluctuation of the adaptive weights around their mean value during learning. It quantifies the degree of dynamical variability in the weight evolution and provides an indication of the balance between plasticity and stability within the system \cite{pmlr-v202-lyle23b}. In the system, the weight variance corresponding to  the $\eta$ values; $[10^{-5}, 10^{-4}, 10^{-3}, 10^{-2}]$ are; $[0.0078, 0.0114, 0.0259, 1.8807]$. By linear interpolation between the minimum and maximum bounds of the $\eta$ and weight variance parameters, 10 values for each parameter were obtained.
The relationship between weight variance and learning rate is shown in Figure \ref{fig:weight variance}, illustrating the stability–plasticity tradeoff in the proposed RNPDN system. As the learning rate increases, the weight variance grows monotonically, indicating a transition from stable to highly plastic system behaviour. For low learning rates, the system exhibits minimal weight fluctuations, reflecting a stable regime in which previously learned states are preserved but adaptability to new inputs is limited. In contrast, higher learning rates result in significantly increased weight variance, corresponding to a highly plastic regime where the system rapidly adapts to new inputs but at the cost of reduced stability and potential loss of previously stored information. This behaviour highlights the inherent tradeoff between stability and plasticity, where increasing the learning rate enhances responsiveness and learning capability while simultaneously introducing variability that can destabilize stored representations. The nonlinear increase in variance further suggests that beyond a certain threshold, small increases in learning rate lead to disproportionately large fluctuations, marking the onset of instability.

\subsection{Hardware-Faithful Learning Dynamics in RNPDN}
\begin{figure}[!htbp]
    \centering
    \includegraphics[width=0.45\textwidth]{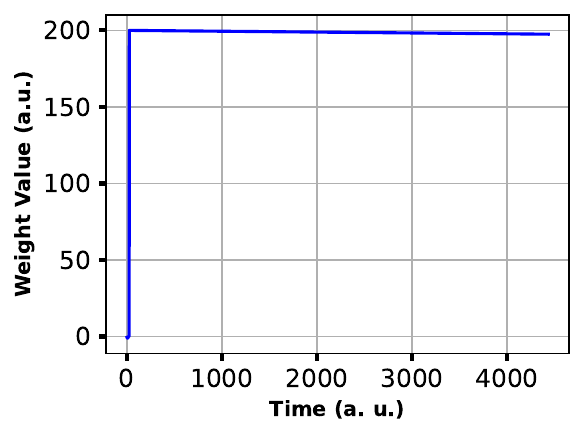}
    \caption{Demonstration of weight saturation as not being a numerical artifact. Setting a reasonable maximum weight value is essential to avoid nonphysical divergence or unstable learning as shown}
    \label{fig:hardware faithful_weight saturation}
\end{figure}

In this subsection, an effort is made to demonstrate that the RNPDN is not just inspired by hardware, but constrained by it. In the context of this system, hardware-faithful means the learning and memory behaviour of the RNPDN arises solely from physical constraints that are directly implementable in photonic hardware, without algorithmic abstractions. In nonlinear photonic systems, the optical response does not increase indefinitely with input intensity but instead exhibits saturation behaviour, leading to a bounded refractive index change and absorption response \cite{rao2022saturation, kir2003absorption}. In RNPDN, weight saturation can mean finite refractive index change/absorption, or finite carrier density or phase-change contrast. In Figure~\ref{fig:weight_trajectory}, the maximum weight value is set to $3.0 (a. u.)$, and this enabled memory to be formed as shown. In Figure~\ref{fig:hardware faithful_weight saturation}, the maximum weight value is set to $10^{200} (a. u.)$ and this resulted in nonphysical divergence as shown. This demonstrates that weight saturation is not a numerical artifact; it is essential for stability and mirrors physical limits of photonic devices.

\begin{figure}[!htbp]
    \centering
    \includegraphics[width=0.45\textwidth]{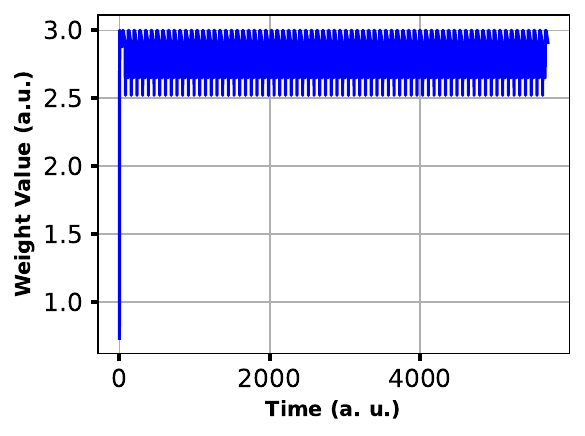}
    \caption{Demonstration that memory cannot be formed when hysteretic component (which induces bistable memory) of RNPDN is put off.}
    \label{fig:hardware faithful_hysteretic switchin}
\end{figure}

Again, the hysteretic component of the RNPDN was switched off and Figure~\ref{fig:hardware faithful_hysteretic switchin} demonstrates that memory cannot be formed when hysteresis, which induces bistable memory is put off. This shows that memory formation in RNPDN arises from hysteretic switching, not from algorithmic state storage.

These proofs reveal that, RNPDN is not an algorithm adapted to hardware, it is a learning system whose behaviour is dictated by hardware physics. Without saturation, the system becomes nonphysical; energy grows unbounded, memory states lose meaning, and hysteresis thresholds stop mattering. Stable learning and memory in RNPDN therefore require finite, hardware-imposed weight saturation.

\section{Demonstration of Reproducibility and Consistency of the RNPDN Framework}
\begin{figure}[!htbp]
    \centering
    \includegraphics[width=0.45\textwidth]{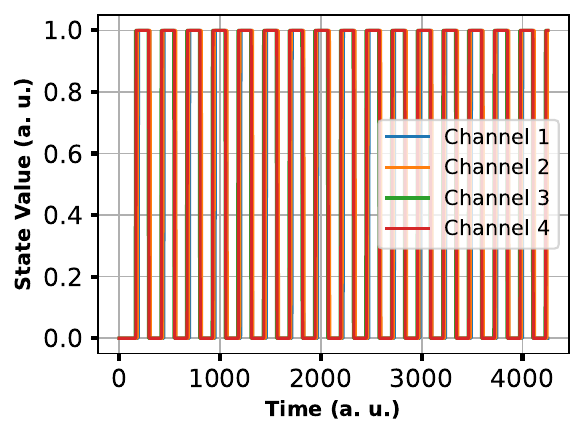}
    \caption{Multi-channel operation of the RNPDN. Four independent channels $S_1 - S_4$ exhibit identical state trajectories, demonstrating reproducible nonlinear dynamics, bistable switching, and memory formation across all channels. The uniformity of the responses confirms the scalability of the approach to larger channel arrays.}
    \label{fig:multi state trajectory}
\end{figure}

\begin{figure}[!htbp]
    \centering
    \includegraphics[width=0.45\textwidth]{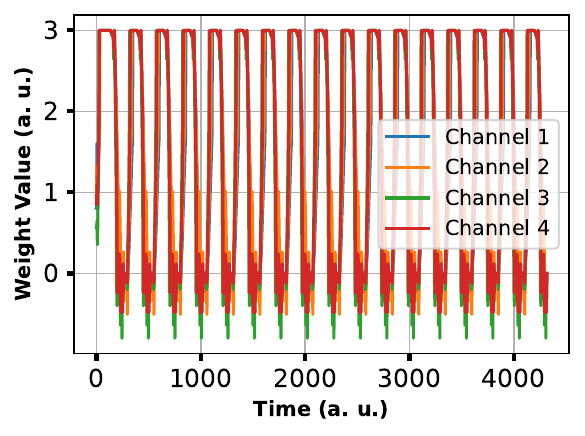}
    \caption{Scalable photonic memory array of RNPDN. All four weight channels $W_1 - W_4$ exhibit identical dynamics: initial plasticity, long-term memory retention, physical saturation and relaxation (decay). This uniformity across channels validates the hardware-faithful scalability of the approach to photonic neuromorphic systems.}
    \label{fig:multi weight trajectory}
\end{figure}

Having established the fundamental dynamics of a single RNPDN unit, the proposed framework was replicated across four identical channels to examine the consistency of its dynamical behaviour. Figures~\ref{fig:multi state trajectory} and ~\ref{fig:multi weight trajectory} show the corresponding state and weight trajectories, respectively. Because all channels were initialized with identical parameters, the resulting trajectories are also identical, as expected for a deterministic dynamical system. Although this result should not be interpreted as a demonstration of large-scale scalability, it confirms that the proposed dynamical model is reproducible across multiple replicated channels and that no additional numerical artefacts arise from parallel instantiation of the framework. This consistency provides an initial indication that the underlying learning dynamics can be replicated across multiple processing units. Future work will investigate true multi-channel operation by introducing channel-specific inputs, parameter variations, coupling between channels, and hardware-induced variability to evaluate robustness, scalability, and parallel processing capabilities under realistic operating conditions.

\subsection{Demonstration of In-situ learning}
\begin{figure}[!htbp]
    \centering
    \includegraphics[width=0.45\textwidth]{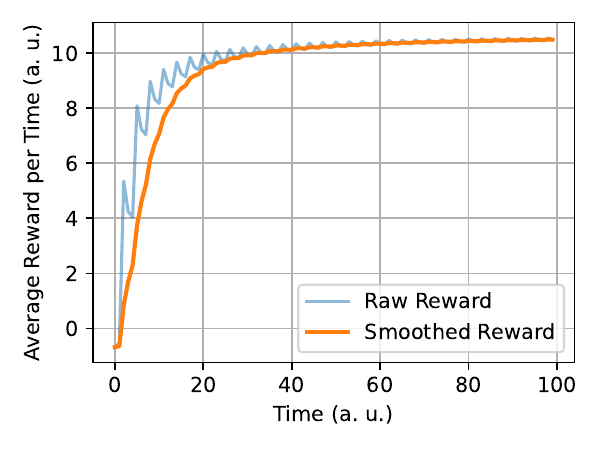}
    \caption{Time evolution of the reward signal during In-situ learning. Fast rise implies the learning phase, oscillations imply nonlinearity or system dynamics and plateau implies convergence. This demonstrates the performance improvement over time.}
    \label{fig:In-situ learning_reward}
\end{figure}

\begin{figure}[!htbp]
    \centering
    \includegraphics[width=0.45\textwidth]{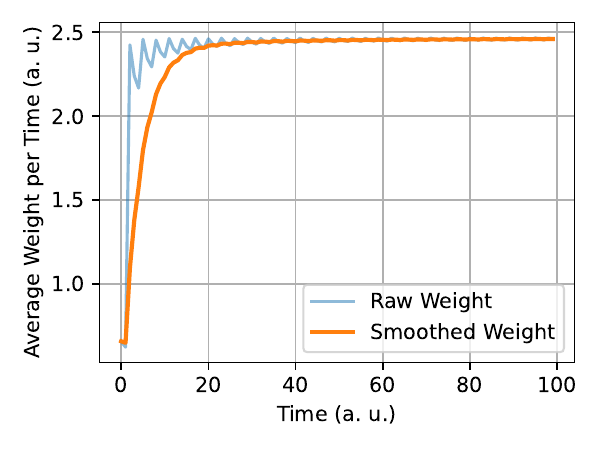}
    \caption{Time evolution of weight during In-situ learning. There is rapid adaptation initially, and then there is convergence. The small residual oscillations suggest internal state adaptation and stabilization.}
    \label{fig:In-situ learning_weight}
\end{figure}

\begin{figure}[!htbp]
    \centering
    \includegraphics[width=0.45\textwidth]{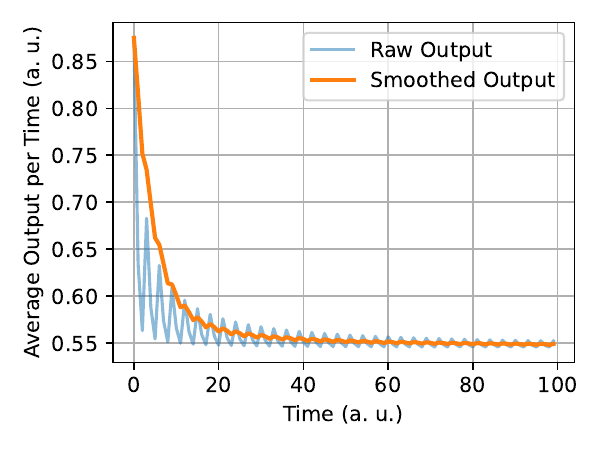}
    \caption{Time evolution of system output. There is high initial value ($> 0.85$), rapid drop, and gradual convergence ($\approx 0.55$). This shows that output settles to a functional operating point.}
    \label{fig:In-situ learning_output}
\end{figure}

In this section, in-situ learning is demonstrated through the coupled evolution of system reward, internal weights and system output, where performance improves over time while the underlying parameters adapt dynamically under a local update rule, eventually converging to a stable operating regime. The time evolution of the reward signal shows initial fast rise, indicating the learning phase, followed by oscillations, which suggests nonlinearity or system dynamics and then plateau which signals convergence. Together, these reveal performance improvement over time. The weight evolution also shows rapid adaptation initially followed by smooth convergence with small residual oscillations. These also reveal internal state adaptation and stabilization over time. The system output evolution also shows similar pattern, confirming system state adaptation and convergence over time. Figure~\ref{fig:In-situ learning_reward} shows the graph of the time evolution of the reward signal, Figure~\ref{fig:In-situ learning_weight} shows that of the weight evolution, and Figure~\ref{fig:In-situ learning_output} shows the system output evolution.

Together, these plots establish that learning occurs during operation and not offline training, internal parameters evolve dynamically (weights are not fixed), and performance improves as a result of increases in reward. It is important to note that convergence of the learning process does not require monotonic increase or strict convergence of the weights. Prior work has shown that neural network weights and outputs may not show monotonic increase. Instead, the key requirement is the convergence of system dynamics and task performance, which is observed in the proposed system \cite{zhang2022neural, matsuoka1992stability, fahey2026linear}. Again, system output does not necessarily need to increase, but rather, it needs to stabilize to a meaningful value \cite{kozachkov2020achieving}, which is what is being observed in Figure~\ref{fig:In-situ learning_output}. Importantly, learning does not mean everything increases, instead; reward should increase, output should converge and weights should stabilize. This suggests that the system is self-regulating its response, which is consistent with nonlinear photonic systems as well  as driven–dissipative dynamics.

\section{Discussion}
The simulation results demonstrate that the proposed RNPDN framework exhibits driven–dissipative learning dynamics, nonlinear state evolution, fading memory, stability–plasticity tradeoffs, and in-situ weight adaptation within the adopted phenomenological model. These observations suggest that such dynamics may provide a physically motivated basis for adaptive computation in photonic systems. However, this work is not intended to provide a comprehensive quantitative comparison with conventional RC or other photonic neuromorphic architectures. Rather, it is presented as an early proof-of-concept framework that demonstrates the feasibility of exploiting driven–dissipative photonic dynamics as a physical substrate for adaptive computation. The primary objective is to establish the underlying concept and its physical motivation, thereby providing a foundation for future studies involving device-level implementations, systematic benchmarking, and experimental validation.
Within this conceptual framework, the proposed approach differs from conventional RC in that the internal states and effective weights are updated continuously through local learning dynamics, rather than relying solely on the optimization of a separate readout layer. While conventional RC primarily exploits fading memory in a fixed dynamical substrate, the RNPDN framework explores the possibility of combining fading memory with persistent nonlinear state evolution through local adaptive dynamics. These observations are intended to illustrate the qualitative behaviour of the proposed framework rather than to establish superiority over existing approaches.
The simulated stability–plasticity tradeoff further suggests that the interaction between nonlinear gain, dissipation, and local weight adaptation can produce operating regimes that balance information retention with adaptability. Although demonstrated using a simplified phenomenological model, this behaviour is consistent with the competing processes found in driven–dissipative photonic systems.
Overall, the presented results provide an initial proof-of-concept demonstration that driven–dissipative photonic dynamics can serve as a physically motivated framework for adaptive learning. While further theoretical development, quantitative benchmarking, device-level modelling, and experimental validation are required, the proposed RNPDN framework establishes a conceptual foundation for future investigations of integrated photonic neuromorphic systems.

\section{Conclusion}
This work introduced the RNPDN framework and presented an early proof-of-concept demonstration, through a unified driven–dissipative model, of physical learning rules, nonlinear dynamics, memory formation, fading memory, stability–plasticity tradeoffs, and in-situ learning. The presented results illustrate the feasibility of embedding adaptive learning within the system dynamics using physically motivated local update rules, thereby establishing a conceptual framework for photonic neuromorphic computing. While the current study is based on a simplified phenomenological model, it provides a foundation for future investigations involving device-level modelling, quantitative benchmarking, and experimental validation in integrated photonic hardware.

\section{Data Availability Statement}
The data supporting the findings of this study are available from the author upon reasonable request, subject to reasonable research and intellectual property considerations.

\section*{Acknowledgment}
I sincerely thank my parents for their unwavering prayers, encouragement, and support throughout this research journey. I also gratefully acknowledge Dr. Peter David Girouard for providing me with a strong foundation in academic research and for his guidance in shaping my research career.

\ifCLASSOPTIONcaptionsoff
  \newpage
\fi

\bibliographystyle{IEEEtran}
\bibliography{references}
\end{document}